\documentclass[prx,twocolumn,showpacs,preprintnumbers,amsmath,amssymb,citeautoscript,superscriptaddress]{revtex4-1}
\usepackage{graphicx}
\usepackage{amssymb, amsfonts, amsmath}
\usepackage{here}
\usepackage{dcolumn}
\usepackage{bm}
\usepackage{color}
\usepackage{here}

\newlength{\figwidth}
\newcommand{\tc}{$T_{{\rm c}}$}

\begin{document}

\title{Divergent nematic susceptibility near the pseudogap critical point in a cuprate superconductor}

\author{K. Ishida}
\author{S. Hosoi}
\thanks{Present address: Department of Materials Engineering Science, Osaka University, Toyonaka 560-8531, Japan.}
\affiliation{Department of Advanced Materials Science, University of Tokyo, Kashiwa, Chiba 277-8561, Japan}
\author{Y. Teramoto}
\author{T. Usui}
\affiliation{Graduate School of Science and Technology, Hirosaki University, Hirosaki, Aomori 036-8561, Japan}
\author{Y. Mizukami}
\affiliation{Department of Advanced Materials Science, University of Tokyo, Kashiwa, Chiba 277-8561, Japan}
\author{K. Itaka}
\affiliation{Institute of Regional Innovation, Hirosaki University, Aomori, Aomori 030-0813, Japan}
\author{Y. Matsuda}
\affiliation{Department of Physics, Kyoto University, Sakyo-ku, Kyoto 606-8502, Japan}
\author{T. Watanabe}
\affiliation{Graduate School of Science and Technology, Hirosaki University, Hirosaki, Aomori 036-8561, Japan}
\author{T. Shibauchi}
\affiliation{Department of Advanced Materials Science, University of Tokyo, Kashiwa, Chiba 277-8561, Japan}
\date{\today}

\begin{abstract}
{
Superconductivity is a quantum phenomenon caused by bound pairs of electrons. In diverse families of strongly correlated electron systems, the electron pairs are not bound together by phonon exchange but instead by some other kind of bosonic fluctuations. In these systems, superconductivity is often found near a magnetic quantum critical point (QCP) where a magnetic phase vanishes in the zero-temperature limit. Moreover, the maximum of superconducting transition temperature $T_{\rm c}$ frequently locates near the magnetic QCP, suggesting that the proliferation of critical spin fluctuations emanating from the QCP plays an important role in Cooper pairing. In cuprate superconductors, however, the superconducting dome is usually separated from the antiferromagnetic phase and $T_{\rm c}$ attains its maximum value near the verge of enigmatic pseudogap state that appears below doping-dependent temperature $T^*$ \cite{Keimer2015}. Thus a clue to the pairing mechanism resides in the pseudogap and associated anomalous transport properties. Recent experiments suggested a phase transition at $T^*$ \cite{Shekhter2013,Zhao2017,Sato2017,Badoux2016,Ramshaw2015,Michon2018,Fujita2014}, yet, most importantly, relevant fluctuations associated with the pseudogap have not been identified. Here we report on direct observations of enhanced nematic fluctuations in (Bi,Pb)$_2$Sr$_2$CaCu$_2$O$_{8+\delta}$ by elastoresistance measurements \cite{Chu2012,Kuo2016}, which couple to twofold in-plane electronic anisotropy, i.e. electronic nematicity. The nematic susceptibility shows Curie-Weiss-like temperature dependence above $T^*$, and an anomaly at $T^*$ evidences a second-order transition with broken rotational symmetry. Near the pseudogap end point, where $T_{\rm c}$ is not far from its peak in the superconducting dome, nematic susceptibility becomes singular and divergent, indicating the presence of a nematic QCP. This signifies quantum critical fluctuations of a nematic order, which has emerging links to the high-$T_{\rm c}$ superconductivity and strange metallic behaviours in cuprates.
}

\end{abstract} 

\maketitle
The pseudogap is an unusual depletion of electronic density of states at low energies in the normal state above $T_{\rm c}$, ubiquitously observed in underdoped cuprates. One of the central unsolved issues in high-$T_{\rm c}$ cuprates is the relationship between the pseudogap phenomenon and superconductivity. Widely discussed possibilities are that the pseudogap is a continuous crossover related to preformed superconducting pairs or a distinct ordered phase coexisting with superconductivity \cite{Keimer2015}. Although conventional measurements including transport and specific heat detected gradual changes at the onset of the pseudogap, recent experiments using sensitive measurement techniques, such as resonant ultrasound \cite{Shekhter2013}, second-harmonic optical response \cite{Zhao2017}, and magnetic torque \cite{Sato2017}, have provided evidence for a thermodynamic phase transition with spontaneous symmetry breaking at $T^*$. The $T^*$ line shows strong doping dependence, and fades at around hole concentration $p\sim0.2$ in the slightly overdoped region. Such a feature favors the existence of a quantum critical point (QCP) inside the superconducting dome, whose relevance to the anomalous transport properties including $T$-linear resistivity \cite{Cooper2009} and a drastic change in the Hall effect \cite{Badoux2016} has been actively discussed. Near the QCP, it is expected that quantum-critical fluctuations are strongly enhanced, which can be a candidate of the glue for superconducting pairing. However, although the strong mass enhancement near the QCP has been reported from quantum oscillations experiments \cite{Ramshaw2015} and recent specific heat measurements at low temperatures \cite{Michon2018}, no direct observation of associated quantum-critical fluctuations has been made. Most importantly, a key question remains unanswered as to what kind of fluctuations develop at the pseudogap QCP. Here we report observation of divergent nematic fluctuations near the pseudogap QCP in a cuprate superconductor.  

If the system undergoes a phase transition involving a symmetry breaking, it is widely acknowledged that measuring the susceptibility of the system is very useful to understand the nature of the ordered state. In case of electronic nematic ordering with broken rotational symmetry, the nematic susceptibility that describes the electronic response to a uniaxial strain
is expected to be enhanced toward the transition. To determine the nematic susceptibility, we use elastoresistance, the measurement of the strain-induced change in the electrical resistivity using a piezoelectric device, which has been recently developed as a powerful probe for electronic nematic fluctuations in iron-based superconductors \cite{Chu2012,Kuo2016}. In the pseudogap state, several  experiments such as neutron scattering \cite{Hinkov2008}, 
anisotropic Nernst effect \cite{Daou2010}, and magnetic torque measurements \cite{Sato2017} have revealed an enhancement of in-plane electronic anisotropy below $T^*$. However, these studies have been based on measurements in YBa$_2$Cu$_3$O$_{6+x}$ (YBCO) with the Cu-O chain structure along the $b$ axis. It has been argued that the Cu-O chains help to promote a preferred direction of nematic domains \cite{Daou2010,Sato2017}, which enables to observe the anisotropy in bulk measurements. In contrast, here we focus on the nematic fluctuations above $T^*$, where domain formation is not an issue, and we chose Bi$_{2-x}$Pb$_x$Sr$_2$CaCu$_2$O$_{8+\delta}$ $(x\approx0.6)$, (Bi,Pb)2212, a system without the Cu-O chain structure. Although Bi$_2$Sr$_2$CaCu$_2$O$_{8+\delta}$ (Bi2212) has an orthorhombic structure due to the super-modulations of Bi-O planes along the orthorhombic $b$ axis, the direction of modulations is 45$^\circ$ tilted from the Cu-O-Cu direction (Fig.\,1a), along which the nematicity is observed in YBCO. Moreover, the modulations can be suppressed by the Pb substitution into Bi site, 
which we confirmed by X-ray diffraction (Fig.\,S1). Thus we compare the results of Bi2212 and (Bi,Pb)2212, from which we conclude that the effect of the off-diagonal modulations is not relevant for the nematic fluctuations along the Cu-O-Cu direction. Another advantage of (Bi,Pb)2212 is that single crystals can be easily cleaved, which is practically important to obtain thin enough samples to transmit the strain over entire thickness from the piezo stacks. 

\begin{figure}
	\begin{center}
		\includegraphics[width=\linewidth]{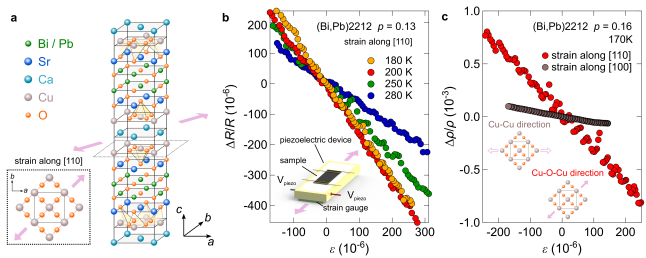}
	\end{center}
		\vspace{-3mm}
	\caption{{\bf Elastoresistance measurements by using the piezoelectric device.} {\bf a}, Crystal structure of (Bi,Pb)2212 in a unit cell. The strain is applied along the [110] direction, parallel to the Cu-O-Cu bonds in the CuO$_2$ plane.  {\bf b},  
		Relative change in the resistance $\Delta R/R$ of a underdoped (Bi,Pb)2212 single crystal at several temperatures as a function of strain induced by the piezoelectric device. Resistance is measured by the conventional four-probe method with current direction parallel to the strain direction. Inset shows the experimental set-up for elastoresistance measurements. The sample (black bars) is glued on top of the piezoelectric device. The amount of the strain can be controlled by applying the voltage to piezo stack and measured by a strain gauge (green) attached underneath the device. {\bf c}, Nematic order parameter $\eta=\Delta \rho/\rho$ as a function of strain along [110] (Cu-O-Cu direction, red circle) and [100] (Cu-Cu direction, brown circle). Here, $\Delta \rho/ \rho$ is derived from elastoresistance data $\Delta R/ R$ (See Methods).
	}
	\label{fig1}
\end{figure}

We measure the change in the resistivity $\eta = \Delta \rho/ \rho$ induced by uniaxial strain which can be controlled by the voltage applied to the piezoelectric device. The quantity $\eta$ is proportional to the resistivity anisotropy, which can be regarded as an electronic nematic order parameter. Since the uniaxial strain $\epsilon$ couples linearly to the nematic order, the nematic susceptibility can be defined as the quantity $\chi_{\rm nem} = \frac{\rm d\eta}{\rm d\epsilon}$ \cite{Chu2012}. Figure\,1b shows typical changes in the resistance $\Delta R/ R$ induced by the uniaxial strain along the Cu-O-Cu direction, which is measured by the strain gauge (Fig.\,1b, inset). The induced changes in the crystal dimensions must be taken into account, and the first-order approximation under an assumption that the sample volume remains constant yields a simple relation $\eta =\Delta \rho/ \rho\approx\Delta R/ R-2\epsilon$
(see Methods). 
Compared to the case when the strain is induced along [100] direction, the response to the strain along [110] direction is much larger (Fig.\,1c). This strong direction dependence ensures that the isotropic $A_{1g}$ component, which admixtures owing to the orthorhombic deformation of piezo stack \cite{Kuo2016}, is quite small compared with the anisotropic nematic component. This also implies that nematic ordering tendency along Cu-O-Cu bond direction is more significant than Cu-Cu direction, consistent with the scanning tunneling spectroscopy (STS) and recent angle-resolved photoemission spectroscopy (ARPES) studies in (Bi,Pb)2212 \cite{Fujita2014,Nakata2018}.  

\begin{figure}
	\begin{center}
		\includegraphics[width=0.6\linewidth]{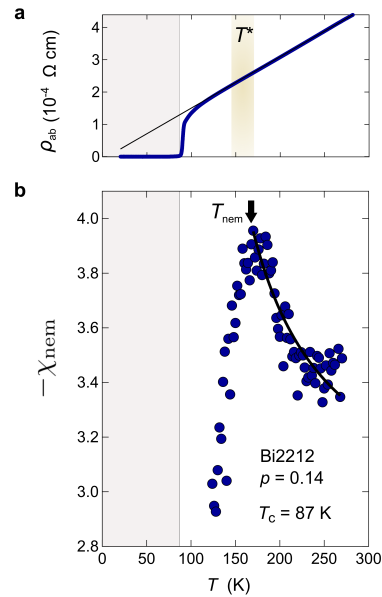}
	\end{center}
		\vspace{-3mm}
	\caption{{\bf Temperature dependence of in-plane resistivity and nematic susceptibility in optimally doped Bi2212.} 
		{\bf a}, Temperature dependence of in-plane resistivity $\rho_{\rm ab}$ in an optimally doped single crystal. Black line is a linear fit to the high temperature data. The pseudogap onset temperature $T^*$ is the deviation point from the $T$-linear dependence (arrow). {\bf b}, Magnitude of nematic susceptibility as a function of temperature in the same sample. The solid curve is a Curie-Weiss fit to the high temperature data (see Eq.\,(1)).  The nematic transition temperature $T_{\rm nem}$ is defined by the deviation from the Curie-Weiss fit, at which $-\chi_{\rm nem}(T)$ shows a kink (arrow). The gray shades indicate the superconducting state below $T_{\rm c}=87$\,K. The two temperatures $T^*$ and $T_{\rm nem}$ coincide well with each other. 
	}
	\label{fig2}
\end{figure}

The temperature dependence of the nematic susceptibility $\chi_{\rm nem}(T)$ for a single crystal of optimally doped Bi2212 with $T_{\rm c}=86$\,K is shown in Fig.\,2 together with the in-plane resistivity data $\rho_{ab}(T)$. The sign of nematic susceptibility is negative, which is the same as in BaFe$_2$As$_2$ \cite{Chu2012,Kuo2016}, but opposite to the case of FeSe-based superconductors \cite{Hosoi2016}. It is more noteworthy that the absolute value of $\chi_{\rm nem}$ is more than an order of magnitude smaller here than in iron-based superconductors. This points to much weaker electron-lattice coupling in this cuprate than in iron-based materials. At high temperatures, $\rho_{ab}(T)$ exhibits the well-known linear-in-$T$ behaviour (Fig.\,2a). In this temperature range, $-\chi_{\rm nem}(T)$ shows a strong increase with decreasing temperature (Fig.\,2b), which can be fitted by a Curie-Weiss law
\begin{equation}
\chi_{\rm nem} = \chi^{0} + \frac{\lambda}{a (T -T_0)} .
\label{eq1}
\end{equation}
Here $\lambda$ and $a$ are constants, and $T_0$ is the Weiss temperature, which represents the electronic nematic transition temperature if there is no electron-lattice coupling \cite{Chu2012,Paul2017}. The electron-lattice coupling increases the actual transition temperature. $\chi^{0}$ is the temperature-independent term associated with the intrinsic piezoresistive response which is not related to the nematic fluctuations. 

Below $T^*\sim 160$\,K, $\rho_{ab}(T)$ deviates gradually from the high-temperature $T$-linear dependence, which has been ascribed to the suppression of quasiparticle scattering rate associated with the pseudogap formation \cite{Watanabe1997}. In the nematic susceptibility data, a clear kink anomaly is observed at a characteristic temperature $T_{\rm nem}$, which is very close to $T^*$. Very similar anomalies in $\chi_{\rm nem}(T)$ have been seen at the tetragonal-orthorhombic structural transition temperature in iron-based superconductors \cite{Chu2012,Kuo2016,Hosoi2016}, below which an electronic nematic phase with strong in-plane anisotropy appears. The similarity strongly implies that an electronic nematic order develops below $T_{\rm nem}$ in this cuprate superconductor. Below the nematic transition temperature $T_{\rm nem}$, it is expected that the nematic domains are formed inside the sample, and thus $\frac{\rm d\eta}{\rm d\epsilon}$ no longer follows the Curie-Weiss law. Although recent Raman scattering experiments could not resolve a clear anomaly at $T^*$ possibly due to the smallness of nematic susceptibility \cite{Auvray2019}, our elastoresistance measurements with a high resolution lead to observe a significant change in the nematic fluctuations at the onset temperature of pseudogap. 

\begin{figure}
	\begin{center}
		\includegraphics[width=\linewidth]{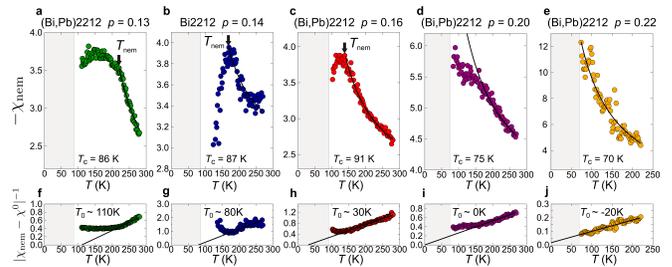}
	\end{center}
	\caption{{\bf Doping dependence of nematic susceptibility in (Bi,Pb)2212.} {\bf a}-{\bf e}, Temperature dependence of the magnitude of nematic susceptibility $-\chi_{\rm nem}$ in (Bi,Pb)2212 for $p=0.13$ with $T_{\rm c}=86$\,K ({\bf a}), 0,16 with $T_{\rm c}=91$\,K ({\bf c}),  0.20 with $T_{\rm c}=75$\,K ({\bf d}),  0.22 with $T_{\rm c}=70$\,K ({\bf e}), and in Bi2212 for $p=0.14$ with $T_{\rm c}=87$\,K ({\bf b}). The solid curves are the Curie-Weiss fits to high-temperature data. The nematic transition temperature $T_{\rm nem}$ is defined as the kink in $-\chi_{\rm nem}(T)$ (arrows in {\bf a}-{\bf c}). {\bf f}-{\bf j}, Inverse susceptibility $|\chi_{\rm nem}-\chi^0|^{-1}$ as a function of temperature. Solid lines are the Curie-Weiss fits, and extrapolations to zero give estimates for the Weiss temperature $T_0$. The gray shades indicate the superconducting state below $T_{\rm c}$.
	}
	\label{fig3}
\end{figure}

As shown in Fig.\,3, the high-temperature Curie-Weiss-like behaviour of the nematic susceptibility has been observed in a wide range of doping in (Bi,Pb)2212, for hole concentration $p$ ranging from underdoped ($p=0.13$) to overdoped ($p=0.22$) regimes, as well as in one slightly underdoped Bi2212 sample ($p=0.14$).  
In the optimally doped and uderdoped regimes (Fig.\,3a-c), $-\chi_{\rm nem}(T)$ shows a deviation from the Curie-Weiss temperature dependence with a kink at $T_{\rm nem}$. 
In contrast, $-\chi_{\rm nem}(T)$ for overdoped samples shows a monotonic increase with decreasing temperature down to $T_{\rm c}$ without showing a kink behaviour (Fig.\,3d,e). 
The Curie-Weiss analysis for high-temperature data yields estimates of Weiss temperature $T_0$ (Fig.\,3f-j), which shows strong doping dependence and changes sign at around $p=0.2$. We note that for this doping $p=0.2$, a downward deviation from Curie-Weiss law can be seen at low temperature. However, similar low-temperature deviations have been reported for several iron-based superconductors in the vicinity of the end point of the nematic transition, where $T_0$ is close to 0\,K \cite{Kuo2016,Hosoi2016}. Thus, the observed deviation is not a signature of the nematic transition as seen in underdoped side, but is likely related to the effect of QCP.

In Fig.\,4 we plot the obtained nematic transition temperature $T_{\rm nem}$ in the phase diagram of Bi2212, which is essentially the same as that of (Bi,Pb)2212 \cite{Hashimoto2014}. It is evident that $T_{\rm nem}$ coincides very well with the onset temperature of pseudogap phenomena $T^*$ reported by ARPES \cite{Hashimoto2014}, superconductor-insulator-superconductor (SIS) tunneling \cite{Hashimoto2014,Ozyuzer2002}, STS \cite{Gomes2007}, and Raman scattering \cite{Sacuto2011} (Fig.\,4a). This coincidence indicates that at the onset of pseudogap, a second-order phase transition with broken rotational symmetry takes place and an electronic nematic order sets in. 

The magnitude of nematic susceptibility changes drastically with doping, as shown in a color plot in Fig.\,4b. It is strongly enhanced near the pseudogap critical point where $T^*\to 0$\,K. If we extrapolate the $T^*$ line linearly to zero, we obtain $p_{\rm c} \sim 0.22$. Concurrently, the sign change of the Weiss temperature $T_0$ occurs very close to $p_{\rm c}$. These results indicate that near the pseudogap QCP, the nematic susceptibility shows a divergent behaviour toward absolute zero, which signifies nematic quantum-critical fluctuations. A nonmagnetic nematic QCP is found for FeSe$_{1-x}$S$_x$, in which a similar divergent nematic susceptibility with enhanced magnitude has been observed \cite{Hosoi2016}. We note that the difference between $T_{\rm nem}$ and $T_0$ is sizable in both (Bi,Pb)2212 and FeSe, suggesting the importance of electron-lattice coupling, while the magnitude of nematic susceptibility is much smaller in (Bi,Pb)2212. According to the recent theory of lattice effects on nematic transition, an important parameter for the electron-lattice coupling is $(T_{\rm nem}-T_0)/E_{\rm F}$, where $E_{\rm F}$ is the Fermi energy \cite{Paul2017}. As the Fermi energy of Bi2212 ($\sim0.3$\,eV) is larger than those of FeSe ($\lesssim0.01$\,eV) \cite{Kasahara2014}, 
even a small electron-lattice coupling can lead to a large difference between $T_{\rm nem}$ and $T_0$, which may explain the experimental results. 

\begin{figure}
	\begin{center}
		\includegraphics[width=1.0\linewidth]{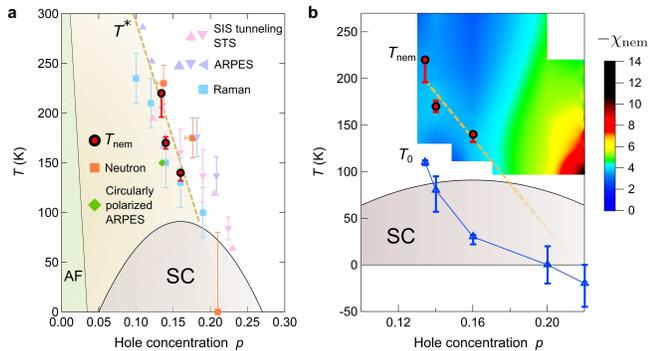}
	\end{center}
	\caption{{\bf Phase diagram and nematic quantum criticality in (Bi,Pb)2212.} 
		{\bf a}, Doping dependence of the nematic transition temperature $T_{\rm nem}$ obtained in the present study for (Bi,Pb)2212 (red circles), compared with the pseudogap temperature reported previously and the dome of superconductivity (SC) for Bi2212. Spectroscopic determined $T^*$ values are from tunnelling experiments (magenta triangles) using SIS junctions \cite{Hashimoto2014,Ozyuzer2002} and STS \cite{Gomes2007}, ARPES (purple triangles) \cite{Hashimoto2014}, and Raman scattering (blue squares) \cite{Sacuto2011}. The temperature below which time reversal symmetry is broken reported by polarized neutron scattering \cite{Almeida-Didry2012} (orange squares) and circularly polarized ARPES \cite{Kaminski2002} (green diamonds) is also plotted. The boundary of the antiferromagnetic (AF) phase is roughly sketched as in the  general phase diagram of hole-doped cuprates \cite{Keimer2015}.
		{\bf b}, Doping dependence of the nematic transition temperature $T_{\rm nem}$ (red circles) and the Weiss temperature $T_0$ (blue triangles) obtained by the Curie-Weiss analysis of the nematic susceptibility data (Fig.\,3). The magnitude of $\chi_{\rm nem}$ is superimposed in the phase diagram by color contours (see the color bar for the scale). The lines are the guides for the eyes. 
	}
	\label{fig4}
\end{figure}

The observed divergent nematic susceptibility continues to grow down to $T_{\rm c}$, below which the system transforms to the superconducting state. Near the quantum critical concentration, $T_{\rm c}$ remains relatively high ($\gtrsim70$\,K). Recent high-field studies indicate that near $p_{\rm c}$ the highest upper critical field is attained and the effective mass is strongly enhanced \cite{Ramshaw2015}. From these results it is tempting to suggest that the quantum-critical nematic fluctuations are linked to high-$T_{\rm c}$ superconductivity in this system. Theoretically, such nematic fluctuations have been considered as a possible source of pairing interactions \cite{Lederer2016}. 
Moreover, the $T$-linear resistivity has been recently observed near the nematic QCP in FeSe$_{1-x}$S$_x$ \cite{Licciardello2019}, where the fundamental question raises as to how the low-carrier-density metals near a ferro-type (${\bf Q}=0$) instability can show non-Fermi liquid inelastic resistivity that requires umklapp processes involving large-${\bf Q}$ momentum transfer. Therefore our results also suggest that the quantum-critical nematic fluctuations may also play a role in the `strange metal' transport behaviour of cuprates. 

Our results provide the experimental evidence that nematic instability is enhanced toward the pseudogap temperature in a cuprate system. 
A recent theory considering many-body vertex corrections suggests that a ferro-type (${\bf Q}=0$) nematic charge modulation develops below $T^*$, and it stabilizes the charge density wave (CDW) with broken translational symmetry (${\bf Q} \neq 0$) at a lower temperature \cite{Kawaguchi2017}. This can explain the enhanced nematic fluctuations near $T^*$ as well as the CDW order observed inside the pseudogap phase \cite{Keimer2015}, but how a ferro-type order can open the pseudogap remains to be resolved. Another question is whether the nematicity is the primary order parameter or some secondary effect of the psuedogap. The possible existence of intra-unit-cell magnetic order in Bi2212, where time-reversal symmetry is broken with translational invariance, has been inferred by the polarized neutron scattering \cite{Almeida-Didry2012} and circularly polarized ARPES measurements \cite{Kaminski2002}. The intra-unit-cell magnetism has been theoretically suggested in the pseudogap phase by considering circulating loop currents in CuO$_2$ plane \cite{Varma2006}, and recent studies suggest that nematicity can coexist with such an exotic order \cite{Fischer2011}. It deserves further experimental investigations to clarify the primary origin of the pseudogap phase.

\section*{Acknowledgements}
We thank fruitful discussion with A. Fujimori, Y. Gallais, B. Keimer, H. Kontani, M. Le Tacon, I. Paul, A. Sacuto, J. Schmalian, L. Taillefer, and C.\,M. Varma. This work was supported by Grants-in-Aid for Scientific Research (Nos. 15H02106, 18H05227), and by a Grant-in-Aid for Scientific Research on Innovative Areas "Quantum Liquid Crystals" (KAKENHI Grant No. JP19H05824) from Japan Society for the Promotion of Science. X-ray diffraction measurements were partly supported by the joint research in the Institute for Solid State Physics, the University of Tokyo. 

\section*{Author contributions}
T.S. conceived the project. K. Ishida, S.H., Y. Mizukami performed elastoresistance measurements. Y.T., T.U. and T.W. synthesized (Bi,Pb)$_2$Sr$_2$CaCu$_2$O$_{8+\delta}$ single crystals. K. Itaka performed X-ray diffraction measurements. All authors discussed the results. K. Ishida, Y. Matsuda, T.S. wrote the paper with inputs from all authors. 



\newpage

\section*{Methods} 
\subsubsection*{\textup{\textbf{Single crystals}}}

Single crystals of Bi$_{2-x}$Pb$_x$Sr$_2$CaCu$_2$O$_{8+\delta}$ with $x=0$ (Bi2212) and $x\approx0.6$ ((Bi,Pb)2212) were grown by the traveling-solvent floating zone method \cite{Watanabe2016}. The doping level is controlled by annealing the crystals under proper conditions of partial oxygen pressure and temperature by considering the equilibrium phase diagram reported by Watanabe {\it et al.} \cite{Watanabe1997}. We determined the hole concentration $p$ using the empirical formula with maximum \tc \ = 89 and 91K for Bi2212 and (Bi,Pb)2212, respectively \cite{Presland1991}.

To check the super-modulation structure of the obtained crystals, we use X-ray diffraction measurements at room temperature in two in-plane geometries that can compare $(h00)$ and $(0k0)$ Bragg peaks as shown in Fig.\,S1. In Bi2212, superlattice satellite peaks can be seen only around the $(0k0)$ Bragg peaks, which is a clear signature of the super-modulations along the $b$ axis. In contrast, such satellite peaks are absent for (Bi,Pb)2212, confirming that the modulation structure is completely suppressed by the Pb substitution.

\setcounter{figure}{0}
\renewcommand{\figurename}{Fig. S\!\!}
\begin{figure}[H]
	\begin{center}
		\includegraphics[width=0.9\linewidth]{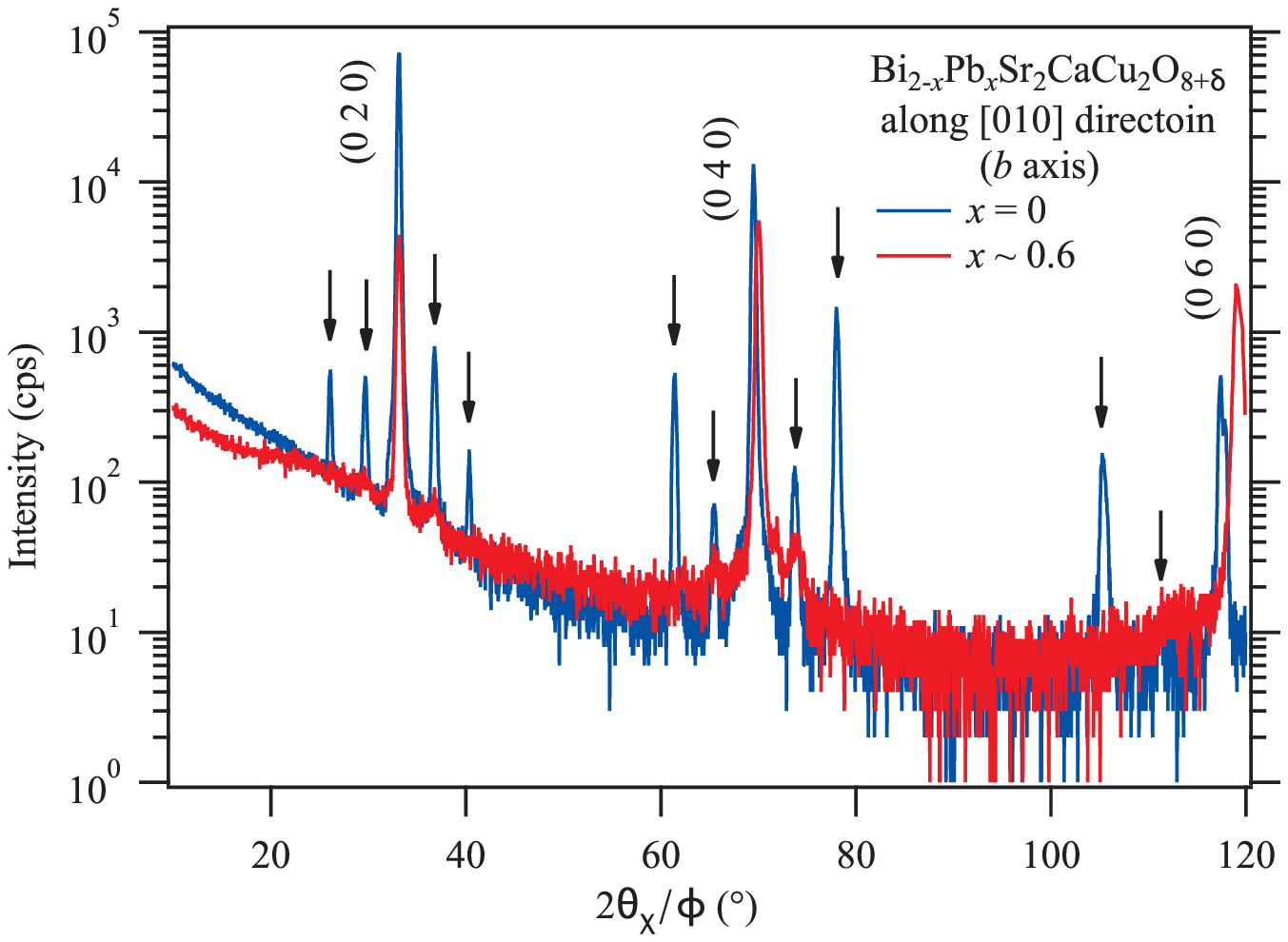} 
	\end{center}
	\caption{
		{\bf In-plane X-ray diffraction for Bi2212 and (Bi,Pb)2212 crystals.} 
		Diffraction patterns for $(0k0)$ Bragg peaks in Bi2212 (blue) and (Bi,Pb)2212 (red) single crystals. The $\phi$ axis is the rotational angle of the sample normal to the $c$ axis, and $\theta_{\chi}$ is the rotational angle of the detector normal to the $c$ axis of the sample. The arrows indicate the satellite peaks due to the super-modulations in Bi2212. }
\end{figure}

\subsubsection*{\textup{\textbf{Elastoresistance measurements}}}
The schematic experimental setup for the elastoresistance measurements is shown in the inset of Fig.\,1b. The samples are cut into rectangular shapes (typical sample size $\sim$ 0.15mm $\times$ 0.70mm $\times$ 0.01mm) along the Cu-O-Cu direction, which is determined by X-ray diffraction. To transmit the strain to samples sufficiently, we cleave the sample to reduce the thickness (along the $c$ axis) down to $\sim 10\,\mu$m. The samples with four contacts for resistivity measurements are then glued on the surface of piezoelectric stack. The strain $\epsilon$ was controlled by applying the voltage to piezo stack and measured by the strain gauge glued to the back surface of the piezo stack. 

To evaluate nematic susceptibility we measure the relative change in the resistance $\Delta R/R$, from which we extract the change in the resistivity $\Delta \rho/\rho$, which can be regarded as an order parameter $\eta$ of the electronic nematic phase, when the strain is induced to the sample.
Here we should consider the change of the sample dimensions due to the physical deformation (geometric effect). As the strain $\epsilon$ is only of the order of $10^{-4}$, higher-order terms of the dimension change are negligibly small. Thus this effect can be expressed as  
\begin{equation*}
\Delta \rho/\rho \approx \Delta R/R - \Delta L/L+\Delta D/D+\Delta W/W, 
\label{eq1}
\end{equation*}
where $\Delta L/L$, $\Delta D/D$ and $\Delta W/W$ are the strain-induced relative changes in the length, depth and width of the sample, respectively. In contrast to the case of Fe pnictides \cite{Kuo2013}, the change of resistance in (Bi,Pb)2212 is not large enough to neglect the geometric factors. To take into account for these geometric effects in the first-order approximation, we assume that the sample volume is preserved during the deformation, which gives 
\begin{equation*}
\Delta L/L+\Delta D/D+\Delta W/W \approx 0.
\label{eq1}
\end{equation*}
Since the $\Delta L/L$ corresponds to the strain $\epsilon$ measured by strain gauge, we can derive a simple relation 
\begin{equation*}
\Delta \rho/ \rho\approx\Delta R/ R-2\epsilon.
\label{eq1}
\end{equation*}

\clearpage

\end{document}